\documentclass[reprint,superscriptaddress,nofootinbib,amsmath,amssymb,aps]{revtex4-1}

\usepackage{dcolumn}
\usepackage{bm}
\usepackage{graphicx}
\usepackage{booktabs}
\usepackage{float}
\usepackage{url}
\usepackage{hyperref}
\usepackage{cleveref}
\usepackage{physics}
\usepackage{braket}

\crefname{figure}{Fig.}{Figs.}
\crefname{equation}{Eq.}{Eqs.}
\crefname{section}{Sec.}{Sec.}
\Crefname{figure}{Figure}{Figures}
\Crefname{equation}{Equation}{Equations}
\Crefname{section}{Section}{Sections}

\DeclareMathOperator*{\argmin}{arg\,min}

\begin{document}
\preprint{APS/123-QED}
\title{Sequential minimal optimization for quantum-classical hybrid algorithms}

\author{Ken M. Nakanishi}
    \email{ken-nakanishi@g.ecc.u-tokyo.ac.jp}
    \affiliation{
        Graduate School of Science,
        The University of Tokyo,
        7-3-1 Hongo, Bunkyo-ku, Tokyo 113-0033, Japan.
    }
\author{Keisuke Fujii}
    \email{fujii.keisuke.2s@kyoto-u.ac.jp}
    \affiliation{
        Graduate School of Science,
        Kyoto University,
        Kitashirakawa Oiwake-cho, Sakyo-ku, Kyoto 606-8302, Japan.
    }
    \affiliation{
        JST, PRESTO, 4-1-8 Honcho, Kawaguchi, Saitama 332-0012, Japan.
    }
\author{Synge Todo}
    \email{wistaria@phys.s.u-tokyo.ac.jp}
    \affiliation{
        Graduate School of Science,
        The University of Tokyo,
        7-3-1 Hongo, Bunkyo-ku, Tokyo 113-0033, Japan.
    }
    \affiliation{
        Institute for Solid State Physics,
        The University of Tokyo,
        5-1-5 Kashiwanoha, Kashiwa, Chiba 277-8581, Japan.
    }

\date{\today}
\begin{abstract}

% There has been a growing interest in the quantum-classical hybrid algorithms for noisy intermediate-scale quantum (NISQ) devices because they are considered impossible to simulate on classical computers.
% These algorithms use a parameterized quantum circuit, and we optimize the parameters of the circuit to minimize the expectation value of a given Hamiltonian by estimating the expectation value iteratively.
% Here 
We propose a sequential minimal optimization method for quantum-classical hybrid algorithms, which converges faster, is robust against statistical error, and is hyperparameter-free.
Specifically, the optimization problem of the parameterized quantum circuits is
divided into solvable subproblems by considering only a subset of the parameters.
In fact, if we choose a single parameter,
the cost function becomes a simple sine curve with period $2\pi$,
and hence we can exactly minimize with respect to the chosen parameter.
Furthermore, even in general cases,
the cost function is given by a simple sum of trigonometric functions with certain periods and hence can be minimized by using a classical computer.
By repeatedly performing this procedure, we can optimize the parameterized quantum circuits so that the cost function becomes as small as possible.
We perform numerical simulations and compare the proposed method with existing gradient-free and gradient-based optimization algorithms.
We find that the proposed method substantially outperforms the existing optimization algorithms and converges to a solution almost independent of the initial choice of the parameters.
This accelerates almost all quantum-classical hybrid algorithms readily and would be a key tool for harnessing near-term quantum devices.

%Taking advantage of the feature of the parameterized quantum circuit effectively, we propose a hyperparameter-free optimization method dedicated to parameterized quantum circuit.
%The numerical simulation results show that our method converges much faster than the existing ones, especially in the condition that statistical error exists.

\end{abstract}

\maketitle
\section{Introduction}\label{sec:intro}

Quantum computing devices with almost a hundred qubits are now within reach in the near future~\cite{Google2018,IBM2017,Intel2019}.
Since they have an unignorable amount of error because of a lack of error correction, they are called noisy intermediate-scale quantum (NISQ) devices.
While the complexity of NISQ devices is limited because of the noise, still their classical simulation is considered to be intractable on classical computers if the fidelity of gates is sufficiently high~\cite{Boixo2018,Bouland2018,Chen2018}.

In order to exploit NISQ devices in useful ways, quantum-classical hybrid algorithms have been proposed.
Quantum-classical hybrid algorithms solve problems with a combination of sampling on low-depth quantum circuits and classical post-processing of the sampled outcomes.
The variational quantum eigensolver (VQE)~\cite{Peruzzo2014,McClean2017,Bauer2016,Kandala2017} is, for example, an attracting algorithm for NISQ devices for finding an approximate ground state of a given Hamiltonian $\mathcal{H}$.
The VQE is now extended in various purposes beyond finding the ground state~\cite{Nakanishi2018,Heya2018}.
The quantum approximate optimization algorithm (QAOA)~\cite{Farhi2014,Farhi2016,Otterbach2017} makes it possible to get approximate solutions for combinatorial optimization problems.
The quantum machine learning~\cite{Biamonte2017}, especially for the near-term devices, has been proposed in various machine learning settings such as supervised learning~\cite{Mitarai2018,Wiebe2014,havlivcek2019supervised}, unsupervised learning~\cite{Lloyd2013}, generative model~\cite{Khoshaman2018}, generative adversarial model~\cite{Dallaire2018}, and so on.

All these major quantum-classical hybrid algorithms have a common structure:
a parameterized quantum circuit and its optimization with respect to an observed cost function.
More precisely, we repeatedly generate an ansatz state $\ket{\psi(\bm{\theta})}$ from a parameterized quantum circuit $U(\bm{\theta})$ on quantum devices and optimize the parameters $\bm{\theta}$ to minimize the expectation value of the given Hermitian operator,
$\braket{\mathcal{H}(\bm{\theta})} = \braket{\psi(\bm{\theta})|\mathcal{H}|\psi(\bm{\theta})}$, on classical computers.
Therefore, the convergence speed of the optimization process is a key factor that determines the overall performance of the algorithms.

Existing works employ 
either gradient-free~\cite{Nelder1965,Powell1964,conn1997convergence}
or gradient-based~\cite{fletcher2013practical,byrd1995limited,kingma2014adam,hestenes1952methods,spall1992multivariate} optimization algorithms.
For example, \citet{Peruzzo2014} and \citet{Ryabinkin2018} use the Nelder-Mead method, which is one of the gradient-free optimization algorithms~\cite{Nelder1965}.
Gradient-based methods are generally more efficient if we can get the gradients directly.
In the context of the neural networks, the backpropagation method, which is an efficient method to calculate the gradients, is one of the most important ingredients to minimize the cost function of the model.
To use gradient-based optimization algorithms for parameterized quantum circuits,
\citet{Li2017} and \citet{Mitarai2018} propose an analytical way to take a partial derivative of parameterized quantum circuits, which allows us to obtain the gradient directly from the expectation value of a Hermitian operator.
However, it is not yet fully explored whether a better optimization algorithm, specially designed for the parameterized quantum circuits, exists, or not.
% For example, the classical variational method [@@@] based on matrix product ansatz for quantum many-body problems employs the structure of the matrix product and minimizes the cost function exactly to the minimum for each parameter iteratively.
For example, in the optimization procedure in the support vector machine~\cite{SVM}, the sequential minimal optimization~\cite{Platt1998} is widely used, in which the cost function is exactly minimized with respect to two parameters at each step by making use of the characteristic structure of the quadratic programming problem.

In this study, we propose an optimization method which is specialized for quantum-classical hybrid algorithms based on parameterized quantum circuits.
The idea is similar to the sequential minimal optimization for support vector machine; the cost function is minimized exactly with respect to certain chosen parameters at each step by using the characteristic structure of the parameterized quantum circuits.
The proposed optimization method has several good properties: hyperparameter-free, faster convergence, less dependence on the initial choice of the parameters, and robust against the statistical error.
To this end, we use the fact that the cost function, as a function of a parameter $\theta$, behaves very simply as a sine curve with period $2\pi$, when the parameterized quantum circuit consists of a unitary gate $\exp(i\theta A)$ being subject to $A^2=I$.
By virtue of this special property, we can determine the angle $\theta$ that provides the exact minimum of the cost function by evaluating the cost function at three independent points with respect to the parameter.
This allows us to update each parameter with fewer measurements and to achieve a smaller value of the cost function.
The update is deterministic if the order of the parameters is provided, and hence the proposed optimization method is hyperparameter-free.

We perform detailed numerical simulations to compare the proposed method with existing optimization algorithms including both of gradient-free and gradient-based methods.
To this end, we develop a benchmark task for which we can purely compare the performance of the optimization algorithms apart from the representation power of the parameterized quantum circuits.
The results show that the proposed optimization method converges extremely faster than the existing ones, especially when the statistical error exists because of the finite number of the samples to estimate an expectation value.
Furthermore, we find that the proposed method converges to the solution almost independently of the choice of the initial parameters, while other methods fail for certain initial parameters.
These results mean that the proposed method readily accelerates almost all quantum-classical hybrid algorithms substantially in a practical situation.
The proposed optimization method should be an inevitable ingredient in the parameter tuning of NISQ devices.

The rest of the paper is organized as follows.
In \cref{sec:method} we first describe the proposed optimization method for a single parameter optimization at each step, where we can find an exact minimum at each step.
Then we further extend it for the case where multiple parameters are updated at each step.
We also explain the case where multiple gates are parameterized by the same parameters.
In \cref{sec:simulation}, we present numerical simulations to compare the proposed method with the existing optimization algorithms.
We perform two tasks.
One is a benchmark of the optimization of the parameterized quantum circuits, where we can achieve the exact solution if an optimization algorithm successfully finds the best parameters regardless of the representation power of the ansatz state.
The other is VQE of the lithium hydride (LiH) molecule.
\Cref{conclusion} is devoted to the conclusion.

\section{Methods}\label{sec:method}

\subsection{Preconditions}\label{sec:preconditions}
Our optimization method requires the following three conditions of the quantum-classical hybrid algorithms with parameterized quantum circuits.
\begin{enumerate}
    \item
    The parameters of the parameterized quantum circuit are independent of each other.
    (This condition can be relaxed by extending the proposed method as we will see in \cref{sec:share_method}.)
    \item
    The parameterized quantum circuit with $J$ parameters
    $U(\bm{\theta})\ \qty(\bm{\theta}:=\qty{\theta_j}_{j=1}^J)$
    is composed only of the two types of gates:
    fixed unitary gates
    (e.g.\ the Hadamard gate and the control-$Z$ gate)
    and
    rotation gates
    \begin{equation}\label{eq:rotation_gate}
        R_j(\theta_j) = \exp(-\frac{i\theta_j}{2} A_j) \ \ (j=1,2,\cdots,J),
    \end{equation}
    where $A_j \  (j=1,2,\cdots,J)$ satisfies the condition
    \begin{equation}\label{eq:rotation_gate_condition}
        A_j^2 = I.
    \end{equation}
    (e.g.\ the $Z$-rotation gate and the $X$-rotation gate.)
    \item
    The cost function which we are going to minimize is written by the (weighted) sum of $K$ expectation values:
    \begin{equation}
        \mathcal{L}(\bm{\theta}) = \sum_{k=1}^K w_k \braket{\varphi_k|U^{\dagger}(\bm{\theta})\mathcal{H}_k U(\bm{\theta})|\varphi_k},
    \end{equation}
    where $\mathcal{H}_k\ (k=1,\cdots,K)$ are Hermitian operators such as Hamiltonian, $\qty{\ket{\varphi_k}}_{k=1}^{K}$ are the input states, and $w_k$ is the weight of the $k$-th term. Hereinafter, we refer to $\mathcal{L}$ as the ``cost function''.
\end{enumerate}
Most quantum-classical hybrid algorithms with parameterized quantum circuits, such as hardware efficient ansatz~\cite{Kandala2017,havlivcek2019supervised}, satisfy these requirements.

\subsection{Our method}\label{sec:our_method}

Here we describe how parameters are updated in the proposed optimization method.
Let $\bm{\theta}^{(n)}$ be the parameters of the circuit after $n$ steps of update.
Then, let $U_j^{(n)}(\theta_j)$ be the parameterized quantum circuit $U(\bm{\theta})$ in which the parameters are fixed to be $\bm{\theta}^{(n)}$ except for the $j$-th parameter $\theta_j$:
\begin{equation}
    U_j^{(n)}(\theta_j)
    := U\qty(\bm{\theta})
    |_{\theta_{j'}=\theta^{(n)}_{j'}\ \qty(\mathrm{for}\ j' \neq j)}.
\end{equation}
Similarly, let $\mathcal{L}_j^{(n)}(\theta_j)$ be the cost function $\mathcal{L}(\bm{\theta})$ with the fixed parameters $\bm{\theta}^{(n)}$ except for the $j$-th parameter $\theta_j$:
\begin{equation}
    \mathcal{L}_j^{(n)}(\theta_j)
    := \mathcal{L}\qty(\bm{\theta})
    |_{\theta_{j'}=\theta^{(n)}_{j'}\ \qty(\mathrm{for}\ j' \neq j)}.
\end{equation}
This cost function $\mathcal{L}_j^{(n)}(\theta_j)$ can be rewritten as a function of $\theta_j$:
\begin{align}
    \mathcal{L}_j^{(n)}(\theta_j)
    &= \sum_{k=1}^K w_k \braket{
    \varphi_k|
    U_j^{(n)\dagger}(\theta_j)\mathcal{H}_k U_j^{(n)}(\theta_j)
    |\varphi_k}
    \notag\\
    &= a_{1j}^{(n)} \cos(\theta_j - a_{2j}^{(n)}) + a_{3j}^{(n)},\label{eq:cost_1p}
\end{align}
where $a_{\ell j}^{(n)} \  (\ell=1,2,3)$ denote constants independent of $\theta_j$.
(See appendix~A for the derivation.)
\Cref{eq:cost_1p} tells us that the relation of $\mathcal{L}_j^{(n)}(\theta_j)$ to $\theta_j$ is just a sine curve with period $2\pi$.
The three constants $a_{1j}^{(n)}, a_{2j}^{(n)}$, and $a_{3j}^{(n)}$
can be determined from the values of the cost function $\mathcal{L}_j^{(n)}(\theta_j)$ evaluated at three independent points of $\theta_j$.
Then we can find the argument $\theta_j$  that minimizes the cost function $\mathcal{L}_j^{(n)}(\theta_j)$, i.e., $\displaystyle \argmin_{\theta_j} \mathcal{L}_j^{(n)}(\theta_j)$.

Using the above feature, we update the parameters as follows:
\begin{enumerate}
    \item Choose an index $j_n \in \qty{1, 2, \cdots, J}$ of the parameters sequentially or randomly.
    \item Estimate $\mathcal{L}_{j_n}^{(n-1)}(\theta_{j_n}^{(n-1)} \pm \frac{\pi}{2})$ from a quantum device.
    Although the arguments do not have to be $\theta_{j_n}^{(n-1)} \pm \frac{\pi}{2}$, the present choice simplifies the optimization at step 3 greatly.
    \item Determine $\theta_{j_n}$ minimizing the cost function $\mathcal{L}_{j_n}^{(n-1)}(\theta_{j_n})$ given by \cref{eq:cost_1p} using $\mathcal{L}_{j_n}^{(n-1)}(\theta_{j_n}^{(n-1)})$, which was obtained in the previous update, and $\mathcal{L}_{j_n}^{(n-1)}(\theta_{j_n}^{(n-1)} \pm \frac{\pi}{2})$. 
    \item Update as follows:
    \begin{align}
        \theta_{j_n}^{(n)} &= \argmin_{\theta_{j_n}} \mathcal{L}_{j_n}^{(n-1)}(\theta_{j_n})\\
        \theta_j^{(n)} &= \theta_j^{(n-1)}
        \quad \qty(\mathrm{for}\ j \neq j_n)\\
        \mathcal{L}_{j_{n+1}}^{(n)}(\theta_{j_{n+1}}^{(n)}) &= \min_{\theta_{j_n}}\mathcal{L}_{j_n}^{(n-1)}(\theta_{j_n}).
    \end{align}
    Note that, the minimum is not estimated directly but calculated from the sine curve.
    Therefore, the statistical error would accumulate.
    To avoid this, the minimum $\mathcal{L}_{j_{n+1}}^{(n)}(\theta_{j_{n+1}}^{(n)})$ should be estimated directly in a certain period.
    \item Repeat step 1, 2, 3, and 4 until $\mathcal{L}(\bm{\theta})$ converges.
\end{enumerate}

\subsection{Generalization of our method}\label{sec:general_method}
In the previous section, the minimization is performed with respect to one parameter at each update.
Here we generalize it for a multi-parameter case.
Let $\bm{\theta}^{(n)}$ be the values of the parameters after $n$ times of update as in~\cref{sec:our_method}.
Then, let $U_M^{(n)}(\qty{\theta_j}_{j \in M})$ be the parameterized quantum circuit $U(\bm{\theta})$ in which the parameters are chosen to be $\bm{\theta}^{(n)}$ except for a set of parameters $\qty{\theta_j}_{j \in M}\ (M \subset \qty{1,\cdots,J})$:
\begin{equation}
    U_M^{(n)}\qty(\qty{\theta_j}_{j \in M}) := U\qty(\bm{\theta})
    |_{\theta_{j'}=\theta^{(n)}_{j'}\ \qty(\mathrm{for}\ j' \notin M)},
\end{equation}
where $M$ is a subset of indices for which the minimization is performed.
Similarly, let $\mathcal{L}_M^{(n)}(\qty{\theta_j}_{j \in M})$ be the cost function $\mathcal{L}(\bm{\theta})$ with the fixed parameters $\bm{\theta}^{(n)}$ except for the parameters $\qty{\theta_j}_{j \in M}\ (M \subset \qty{1,\cdots,J})$:
\begin{equation}
    \mathcal{L}_M^{(n)}\qty(\qty{\theta_j}_{j \in M}) := \mathcal{L}\qty(\bm{\theta})
    |_{\theta_{j'}=\theta^{(n)}_{j'}\ \qty(\mathrm{for}\ j' \notin M)}.
\end{equation}
This cost function $\mathcal{L}_M^{(n)}(\qty{\theta_j}_{j \in M})$ can be rewritten as
\begin{align}
    &\mathcal{L}_M^{(n)}\qty(\qty{\theta_j}_{j \in M})
    \notag\\
    &= \sum_{k=1}^K w_k
    \braket{\varphi_k|
    U_M^{(n)\dagger}\qty(\qty{\theta_j}_{j \in M})
    \mathcal{H}_k
    U_M^{(n)}\qty(\qty{\theta_j}_{j \in M})
    |\varphi_k}
    \notag\\
    &= \bm{b}_M^{(n)} \cdot \qty[ \bigotimes_{j\in M}
    \mqty(\cos\theta_j \\ \sin\theta_j \\ 1)],\label{eq:cost_general}
\end{align}
where $\bigotimes$ denotes the Kronecker product, and $\bm{b}_M^{(n)}$ denotes a $3^{\qty|M|}$-dimensional coefficient vector
with $\qty|M|$ being the number of the elements of $M$.
(See appendix~B for the derivation of \cref{eq:cost_general}.)
Since \cref{eq:cost_general} has $3^{\qty|M|}$ parameters $\bm{b}_M^{(n)}$,
we can determine $\bm{b}_M^{(n)}$ using the values of cost function $\mathcal{L}_M^{(n)}\qty(\qty{\theta_j}_{j \in M})$ evaluated at $3^{\qty|M|}$ independent points of $\qty{\theta_j}_{j \in M}$.
Then we can find the values of $\qty{\theta_j}_{j \in M}$ that minimize the cost function $\mathcal{L}_M^{(n)}\qty(\qty{\theta_j}_{j \in M})$.

The optimization for the multi-parameter update runs as follows:
\begin{enumerate}
    \item Choose a subset $M_n \subset \qty{1, 2, \cdots, J}$ of indices.
    
    % \item Calculate $\mathcal{L}_{M_n}^{(n-1)} \qty(\qty{\theta_j^{(n-1)} + \frac{\pi}{2} \alpha_j}_{j \in M_n})$ using a quantum device,
    % where the set $\alpha \in \qty{0, \pm 1}^{\otimes \qty|M_n|} \setminus \qty{0}^{\otimes \qty|M_n|}$ and its $j$-th element $\alpha_j$ are defined.
    
    \item Estimate $\mathcal{L}_{M_n}^{(n-1)} \qty(\qty{\theta_j^{(n-1)} + \frac{2\pi}{3} \alpha_j}_{j \in M_n})$
    for all $\bm{\alpha} \in \qty{0, \pm 1}^{\otimes \qty|M_n|} \setminus \qty{0}^{\otimes \qty|M_n|}$
    using a quantum device,
    where $\alpha_j$ denotes the $j$-th element of $\bm{\alpha}$.
    Although the above choices of the parameters are not necessary, they have the advantage that the coefficients in \cref{eq:cost_general} can be easily determined by using the discrete Fourier transformation.
    \item Determine $\qty{\theta_j}_{j \in M_n}$ that minimizes the cost function $\mathcal{L}_{M_n}^{(n-1)}\qty(\qty{\theta_j}_{j \in M_n})$ from \cref{eq:cost_general} using
    $\mathcal{L}_{M_n}^{(n-1)}\qty(\qty{\theta_j^{(n-1)} + \frac{2\pi}{3} \alpha_j}_{j \in M_n})$
    for $\bm{\alpha} \in \qty{0, \pm 1}^{\otimes \qty|M_n|}$.
    
    \item Update as follows:
    \begin{gather}
        \qty{\theta_j^{(n)}}_{j \in M_n} 
        = \argmin_{\qty{\theta_j}_{j \in M_n}}
        \mathcal{L}_{M_n}^{(n-1)}\qty(\qty{\theta_j}_{j \in M_n})\\
        \theta_j^{(n)} = \theta_j^{(n-1)}
        \quad \qty(\mathrm{for}\ j \notin M_n)\\
        \mathcal{L}_{M_{n+1}}^{(n)}\qty(\qty{\theta_j^{(n)}}_{j \in M_n})
        = \min_{\qty{\theta_j}_{j \in M_n}}
        \mathcal{L}_{M_n}^{(n-1)}\qty(\qty{\theta_j}_{j \in M_n}).
    \end{gather}
    Similarly to the previous case, 
    $\mathcal{L}_{M_{n+1}}^{(n)}\qty(\qty{\theta_j^{(n)}}_{j \in M_n})$
    should be estimated directly in a certain period to avoid the error accumulation.
    \item Repeat step 1, 2, 3, and 4 until $\mathcal{L}(\bm{\theta})$ converges.
\end{enumerate}

\subsection{Special case of our method}\label{sec:share_method}
In this section, we consider the case where the several rotation gates in the parameterized quantum circuit share the same parameter.
This is the case when the target state has a symmetry, like translation invariance, and hence the ansatz state is also subject to it.
Assume that the parameterized quantum circuit has $J$ parameters $\bm{\theta}=\qty{\theta_j}_{j=1}^J$ which are independent of each other.
Unlike \cref{sec:our_method,sec:general_method},
each $\theta_j$ is used at $S_j$ times in the circuit.
% Let $\bm{\theta}^{(n)}$ be the values of the parameters after $n$ times of updating.
% Then, let $U_j^{(n)}(\theta_j)$ be the parameterized quantum circuit $U(\bm{\theta})$ in which the parameters are chosen to be $\bm{\theta}^{(n)}$ except for the $j$-th parameter $\theta_j$:
% \begin{equation}
%     U_j^{(n)}(\theta_j)
%     := U\qty(\bm{\theta})
%     |_{\theta_{j'}=\theta^{(n)}_{j'}\ \qty(\mathrm{for}\ j' \neq j)}.
% \end{equation}
% Similarly, let $\mathcal{L}_j^{(n)}(\theta_j)$ be the cost function $\mathcal{L}(\bm{\theta})$ with the fixed parameters $\bm{\theta}^{(n)}$ except for the $j$-th parameter $\theta_j$:
% \begin{equation}
%     \mathcal{L}_j^{(n)}(\theta_j)
%     := \mathcal{L}\qty(\bm{\theta})
%     |_{\theta_{j'}=\theta^{(n)}_{j'}\ \qty(\mathrm{for}\ j' \neq j)}.
% \end{equation}
$\bm{\theta}^{(n)}, U_j^{(n)}(\theta_j),$ and $\mathcal{L}_j^{(n)}(\theta_j)$ is defined in the same way as \cref{sec:our_method}.
The cost function $\mathcal{L}_j^{(n)}(\theta_j)$ can be transformed as
\begin{align}
    \mathcal{L}_j^{(n)}(\theta_j)
    &= \sum_{k=1}^K w_k
    \braket{\varphi_k|
    U_j^{(n)\dagger}(\theta_j)\mathcal{H}_k U_j^{(n)}(\theta_j)
    |\varphi_k}
    \notag\\
    &= \sum_{s=1}^{S_j} a_{sj}^{(n)} \cos(s\theta)
    + \sum_{s=1}^{S_j} b_{sj}^{(n)} \sin(s\theta)
    + c_j^{(n)},\label{eq:cost_special}
\end{align}
where $a_{sj}^{(n)}, b_{sj}^{(n)} \  (s=1,\cdots,S_j)$ and $c_j$ denote some constants.
Note that, these constants are independent of $\theta_j$.
(See appendix~C for the derivation.)
Since \cref{eq:cost_special} has $2S_j+1$ parameters, $a_{sj}^{(n)}, b_{sj}^{(n)} \  (s=1,\cdots,S_j)$ and $c_{j}^{(n)}$,
we can determine these constants using the values of cost function $\mathcal{L}_j^{(n)}(\theta_j)$ at $2S_j+1$ different values of $\theta_j$, and then find $\theta_j$ minimizing the cost function $\mathcal{L}_j^{(n)}(\theta_j)$.

% Using this feature, we propose an updating method as follows:
% \begin{enumerate}
%     \item Choose a number $j_n \in \qty{1, 2, \cdots, J}$.
%     \item Estimate $\mathcal{L}_{j_n}^{(n-1)}(\theta_{j_n}^{(n-1)} + \frac{2\pi s}{S_j+1}) \ (s=1,\cdots,S_j)$ using a quantum device.
%     \item Determine $\theta_{j_n}$ minimizing the cost function $\mathcal{L}_{j_n}^{(n-1)}(\theta_{j_n})$ by \cref{eq:cost_special} using  $\mathcal{L}_{j_n}^{(n-1)}(\theta_{j_n}^{(n-1)} + \frac{2\pi s}{S_j+1}) \ (s=0,\cdots,S_j)$. 
%     \item Update as follows:
%     \begin{align}
%         \theta_{j_n}^{(n)} &= \argmin_{\theta_{j_n}} \mathcal{L}_{j_n}^{(n-1)}(\theta_{j_n})\\
%         \theta_j^{(n)} &= \theta_j^{(n-1)}
%         \quad \qty(\mathrm{for}\ j \neq j_n)\\
%         \mathcal{L}_{j_{n+1}}^{(n)}(\theta_{j_{n+1}}^{(n)}) &= \min_{\theta_{j_n}}\mathcal{L}_{j_n}^{(n-1)}(\theta_{j_n}).
%     \end{align}
%     Note that, the value $\mathcal{L}_{j_{n+1}}^{(n)}(\theta_{j_{n+1}}^{(n)})$ had better be recalculated using NISQ device several times in order to avoid accumulating noise.
%     \item Repeat step 1, 2, 3, and 4 until $\mathcal{L}(\bm{\theta})$ converges.
% \end{enumerate}
Using this feature, we propose an updating method which is the same method of \cref{sec:our_method} except for the following points:
\begin{itemize}
    \item Estimate $\mathcal{L}_{j_n}^{(n-1)}(\theta_{j_n}^{(n-1)} + \frac{2\pi s}{2S_j+1}) \ (s=1,\cdots,2S_j)$ using a quantum device.
    Although the above choices of the parameters are not necessary, they have the advantage that the coefficients can be determined easily by using the discrete Fourier transformation.
    \item Determine $\theta_{j_n}$ minimizing the cost function $\mathcal{L}_{j_n}^{(n-1)}(\theta_{j_n})$ [\cref{eq:cost_special}] by using  $\mathcal{L}_{j_n}^{(n-1)}(\theta_{j_n}^{(n-1)} + \frac{2\pi s}{2S_j+1}) \ (s=0,\cdots,2S_j)$. 
\end{itemize}

\section{Numerical Simulation}\label{sec:simulation}

\subsection{Numerical setups}

In this section, we numerically demonstrate the performance of the proposed method via two types of optimization tasks.
One (task~1) is a benchmark task of the parameter optimization problems, which we introduced here to compare the performance of different optimization algorithms.
In this task, we minimize
\begin{align}
    \mathcal{L}(\bm{\theta})
    = - \qty|\bra{0}^{\otimes r}U^\dagger(\bm{\theta}^*)
    U(\bm{\theta})\ket{0}^{\otimes r}|^2,
\end{align}
where $r$ is the number of qubits, and $\bm{\theta}^*$ is randomly chosen from a uniform distribution $[0, 2\pi)$ a priori.
This task is equivalent to VQE whose Hamiltonian is given by 
$\mathcal{H} = U(\bm{\theta}^*)\!\ket{0}^{\otimes r}\bra{0}^{\otimes r}\!U(\bm{\theta^*})$.
However, in our numerical simulations, we sample the output from the quantum circuit $U^{\dag}(\bm{\theta}^*)U(\bm{\theta})|0\rangle ^{\otimes r}$ in the $\{|0\rangle , |1\rangle\}$ basis for each qubit, which can be done even on an actual quantum device. Then we calculate the probability to obtain the outcome zero for all qubits similarly to the estimation of the inner product in \citet{Higgott2018}.
% The method of estimating inner products are proposed by \citet{Higgott2018}.
%    One of the eigenvalues of this Hamiltonian $\mathcal{H}$ is $-1$, and the others are $0$.
The cost function multiplied by $-1$ corresponds 
to the fidelity between two quantum states $U(\bm{\theta})\!\ket{0}^{\otimes r}$ and $U(\bm{\theta}^*)\!\ket{0}^{\otimes r}$.
Since the minimum of the $\mathcal{L}(\bm{\theta})$ is exactly $-1$ and achieved at least when $\bm{\theta} = \bm{\theta}^*$,
we can purely compare the performance of the optimization algorithms apart from the representation power of the parameterized quantum circuits.

The other (task 2) is VQE for the lithium hydride (LiH) molecule at bond distance with four qubits.
% The molecular Hamiltonian $\mathcal{H}_{\mathrm{H_2}}$ is calculated by OpenFermion~\cite{McClean2017} and Psi4~\cite{Parrish2017}.
% We use the STO-3G minimal basis set so that we obtained a 4-qubit Hamiltonian.
% We then map the Hamiltonian onto the space of qubits by the Jordan-Wigner transformation.
The molecular Hamiltonian $\mathcal{H}_{\mathrm{LiH}}$ is the same as the one given in the supplemental material of \citet{Kandala2017}.

We use the method with the single-parameter minimization proposed in~\cref{sec:our_method} at each update.
We re-estimated the cost function $\mathcal{L}_{j_{n+1}}^{(n)}(\theta_{j_{n+1}}^{(n)})$ once in 32 iterations.
We compare the proposed method with five existing optimization algorithms, Powell, Nelder-Mead, conjugate gradient (CG), BFGS, and SPSA methods, implemented in the SciPy library~\cite{SciPy} except for the SPSA method.%
\footnote{
    Actually, we used a slightly modified code of them not to finish the optimization process in the middle.
    This modified code used for our experiment is available at \url{https://github.com/ken-nakanishi/scipy}.
}
The hyperparameters of the SPSA method we used are the same as the default values in Qiskit v0.8.0~\cite{Qiskit}.

The initial values of the circuit parameters were randomly sampled from a uniform distribution $[0, 2\pi)$.
%%%
\begin{figure}[t]
    \centering
    \includegraphics[width=\linewidth]{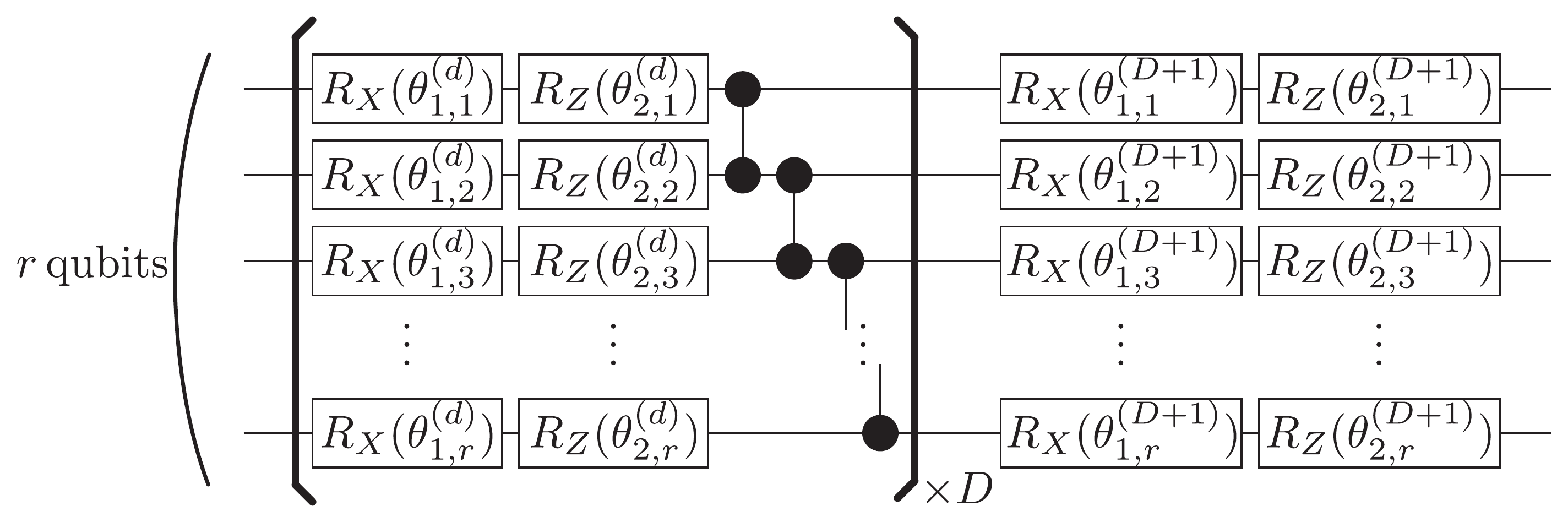}
    \caption{
        The parameterized quantum circuit used in the simulations of \cref{sec:simulation}.
        These parameters $\bm{\theta}$ are optimized to to minimize $\mathcal{L}$.
        $D$ denotes the number of repetition of a circuit in the bracket.
    }
    \label{fig:RX_RZ_CZ_n}
\end{figure}
%%%
In the present numerical simulations, we sampled the outcome 1024 times for the estimation of the cost function (except for the results shown in \cref{fig:res_error}), which determine the amount of the statistical error.
For each simulation, the optimization was run 100 times starting from different initial values of the parameters.
In the following, \textit{step} counts the number of the estimation of the expectation value of the cost function $\mathcal{L}(\bm{\theta})$.

\subsection{Numerical results}
\begin{figure*}[htbp]
    \centering
    \includegraphics[width=\linewidth]{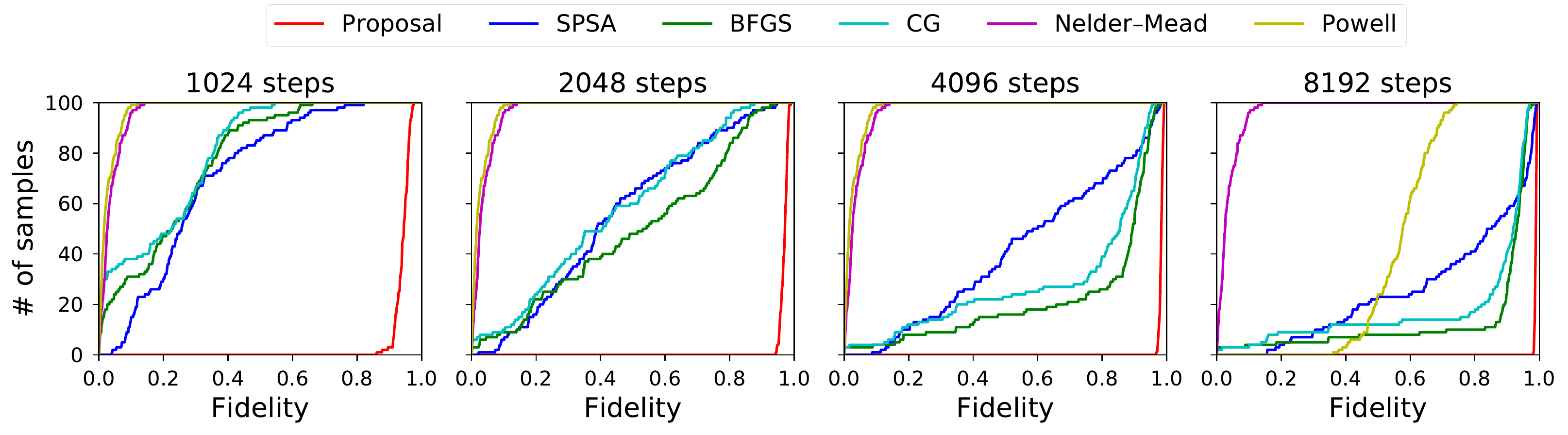}
    \caption{
        Cumulative distribution function of fidelity after 1024, 2048, 4096, and 8192 steps in task~1.
        The proposed method is denoted by the red lines.
        The horizontal axes are the fidelity $\qty|\bra{0}^{\otimes r}U^\dagger(\bm{\theta}^*)U(\bm{\theta})\ket{0}^{\otimes r}|^2$.
        The vertical axis shows the number of samples whose fidelity is under the value of $x$-axis at particular steps.
    }
    \label{fig:res1}
\end{figure*}

In task~1, we set the number of qubits and the depth of the circuit in \cref{fig:RX_RZ_CZ_n}  to be $r=5$ and $D=9$, respectively.
The total number of the parameters is thus 100.
%In this task, the number of qubits is set to 5, the depth $D$ in \cref{fig:RX_RZ_CZ_n} is set to 9, and hence the circuit has 100 parameters.
In \cref{fig:res1},
% we show the fidelity in the horizontal axis for each sample of the initial parameters in vertical axis for each of 1024, 2048, 4096, and 8192 steps from left to right.
the horizontal axes represent the fidelity, and the vertical axes represent the number of samples whose fidelity is under the value of $x$-axis for each of 1024, 2048, 4096, and 8192 steps from left to right.
From \cref{fig:res1}, one can see that our method colored by red converges extremely faster than the other methods we compared.
Specifically, our method achieved fidelity higher than 0.98 after 8192 steps being independent of the initial set of parameters, while the other methods only result in far low fidelity for certain choices of the initial parameters.
\Cref{fig:res1} also shows that the gradient-based methods converge much faster than the gradient-free methods in general.
Regarding the gradient-based methods, BFGS seems to be the best, while SPSA exhibits a better fidelity for some initial parameters.
Furthermore, the Powell method outperforms the Nelder-Mead method when the number of steps is sufficiently large.

\begin{figure*}[htbp]
    \centering
    \includegraphics[width=\linewidth]{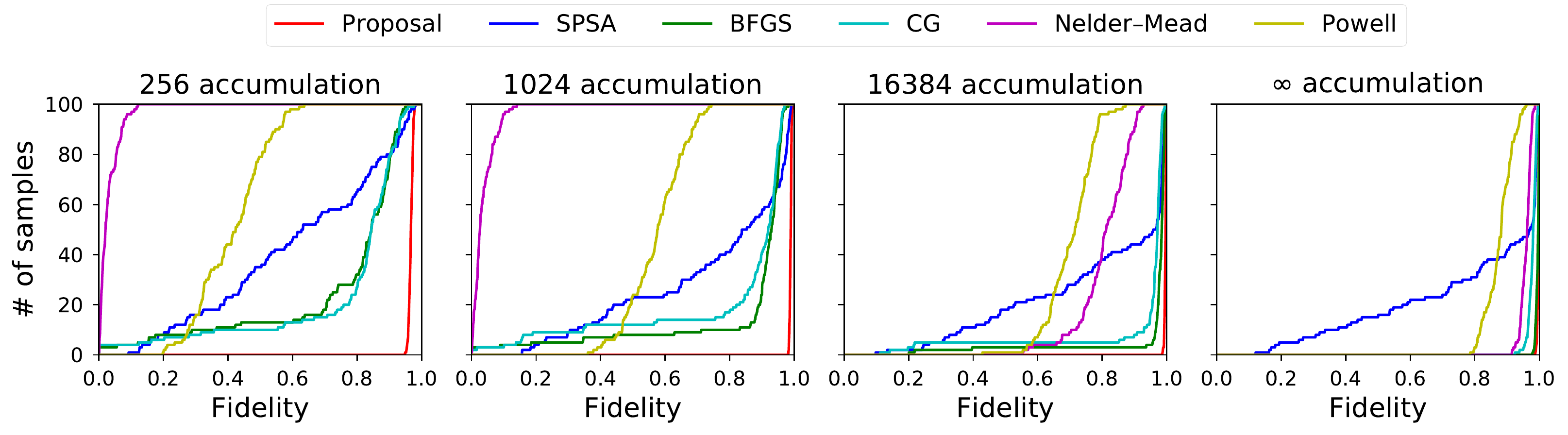}
    \caption{
        Dependence of cumulative distribution function of fidelity on the strength of statistical error.
        The proposed method is denoted by the red lines.
        The horizontal axes are the fidelity $\qty|\bra{0}^{\otimes r}U^\dagger(\bm{\theta}^*)U(\bm{\theta})\ket{0}^{\otimes r}|^2$.
        The vertical axis shows the number of samples whose fidelity is under the value of $x$-axis after 8192 steps.
    }
    \label{fig:res_error}
\end{figure*}

To investigate the statistical-error tolerance of each method,
the number of samples to estimate the cost function is changed.
In \cref{fig:res_error}, we show the results after 8192 steps, for each of 256, 1024, 16384, and $\infty$ outcomes from left to right.
Here $\infty$ means that the cost function is directly calculated from the inner product.
In the limit of the larger number of accumulations, Nelder-Mead, CG, BFGS, and the proposed method achieve high fidelity with almost no dependence on the initial choice of the parameters.
Specifically, BFGS and the proposed method both result in fidelity close to unit.
However, if the statistical error becomes larger with fewer accumulations, the advantage of the proposed method gets larger.
Notably, the proposed method can achieve fidelity above 0.9x with only 256 accumulations for almost all choices of the initial parameters.
Thus we conclude that the proposed method is robust against the statistical error.

\begin{figure*}[htbp]
    \centering
    \includegraphics[width=\linewidth]{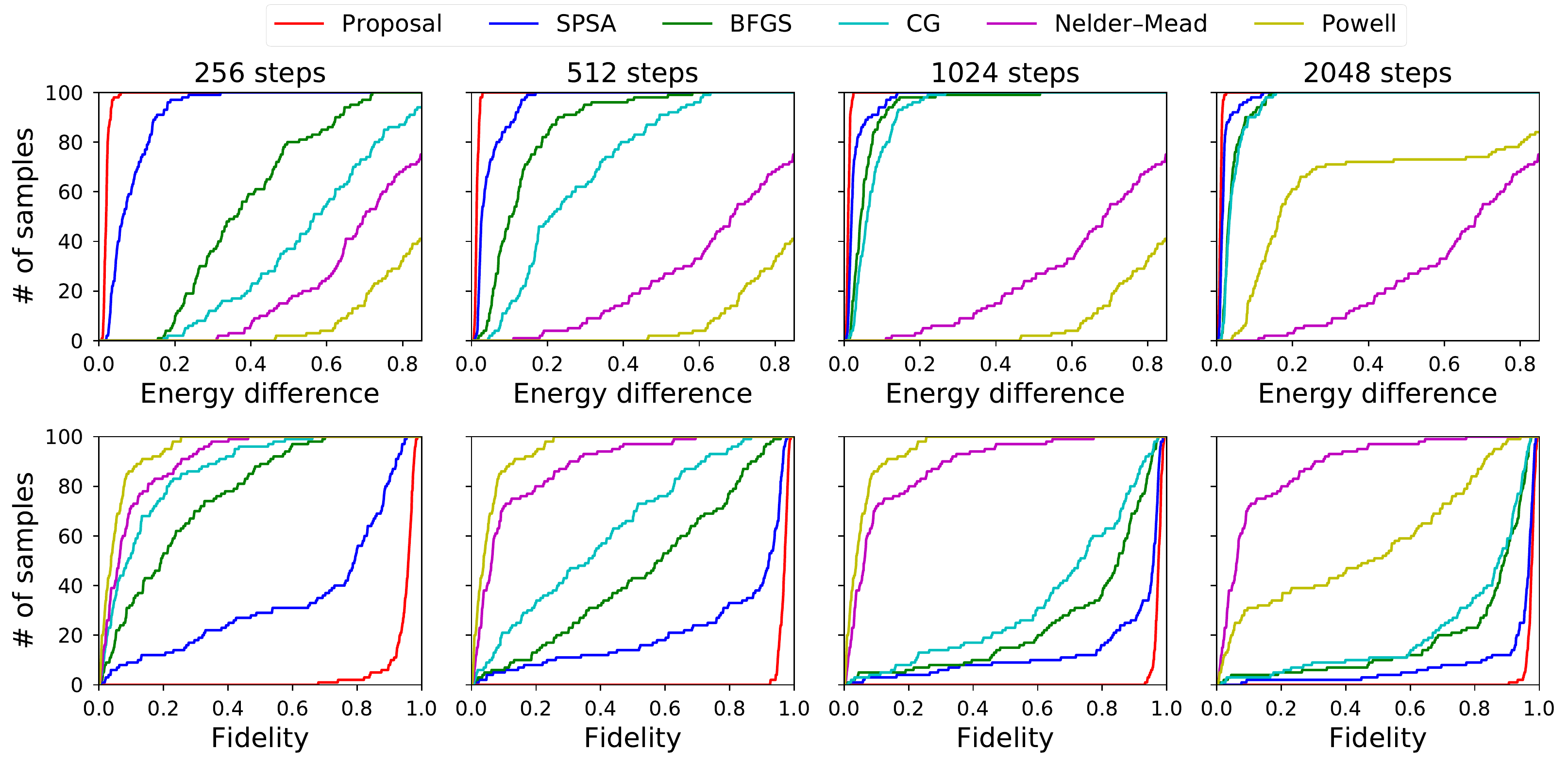}
    \caption{
        Cumulative distribution function of energy difference and fidelity in VQE for the ground state of the Hamiltonian of lithium hydride molecule $\mathrm{LiH}$ (task~2).
        The proposed method is denoted by the red lines.
        The horizontal axes are the energy difference/fidelity between the calculated ground state and the true ground state.
        The vertical axes show the number of samples whose energy difference/fidelity is under the value of $x$-axis at particular steps.
    }
    \label{fig:res2}
\end{figure*}

In task 2, the depth $D$ in \cref{fig:RX_RZ_CZ_n} is set to 4, and hence the circuit has 40 parameters.
We show the results of VQE of the lithium hydride molecule in \cref{fig:res2}.
The horizontal axes represent the energy difference (for top four figures) and fidelity (for bottom four figures) between the solution of VQE and the true ground state.
The vertical axes represent the choices of the initial parameters, which are sorted by the values of each of energy difference and fidelity.
From \cref{fig:res2}, one can see that our method gets closer to the true ground state much faster than the other methods we compared.
Especially, our method achieved fidelity higher than 0.95, which is considered the close-to-limit of the representation power of the prepared parameterized quantum circuits, after only 512 steps against almost all initial set of parameters, while the other methods only result in far low fidelity for certain choices of the initial parameters.

Let us finally argue why the proposed method outperforms the other existing optimization algorithms.
The possible weak point of BFGS and CG methods is that these methods require the gradients of all the parameters.
They are good methods when we can calculate all gradients at once, such as neural-network models.
In the quantum-classical hybrid algorithm, however, we need to estimate the cost function twice as many as the number of the parameters to get all gradients of the parameter.
This could be a disadvantage of these methods, since the step counts not the number of updates but the number of the estimation of the cost function in total.
% , and this can slow down these methods.
In SPSA method, we use the difference of the two cost functions $\frac{\mathcal{L}(\bm{\theta} + \delta\bm{\theta}) - \mathcal{L}(\bm{\theta} - \delta\bm{\theta})}{2\qty|\delta\bm{\theta}|}$ instead of the gradient.
This method cannot avoid the additional noise on the gradient, and this can interrupt the fast convergence.
In task~2, the SPSA method is better results than in task~1.
It is considered that the gradients of the parameters in task~2 is farther apart of zero than ones in task~1 and then can be estimated better.

\section{Conclusion}\label{conclusion}

In this work, we proposed an efficient optimization method for quantum-classical hybrid algorithms using parameterized quantum circuits.
In most quantum-classical hybrid algorithms with parameterized quantum circuits,
the relation of the cost function to each parameter of the circuit is just a sine curve with period $2\pi$, on which our proposed method is based.
To make a good use of the above property, we divide the optimization problem of the parameterized quantum circuits into solvable subproblems by considering only a subset of the parameters.
By numerical simulations, we demonstrate that our method converges to a better solution much faster than the existing ones, especially in the presence of a large statistical error.
The proposed method is expected to have robustness not only to the statistical error but also to the noise in the NISQ devices due to the lack of the error correction, because the relation of the cost function to each parameter of the circuit would be robust against noise.
This property of the present method would enable us to conduct parameterized-quantum-circuit-based variational algorithms on real quantum devices.
We believe that this work drastically accelerates the quantum-classical hybrid algorithms and makes them practical in a realistic situation.
Furthermore, the proposed method itself can be applied not only for optimizing parameterized quantum circuits but also for classical variational method.
For example, the proposed method might be applied for the optimization of multi-scale entanglement renormalization ansatz (MERA)~\cite{vidal2008class} by using variational unitary matrices inspired by parameterized quantum circuits.
Neural networks composed of rotation matrices might also be considered as the another candidate.
% In addition, we used fixed accumulate number for each job.
% You had better increase the accumulate number as the iteration of your job.
\section*{Acknowledgment}
KMN thanks IPA for its support through MITOU Target program.
The computation in this work has been partly done using the facility of the Supercomputer Center, Institute for Solid State Physics, The University of Tokyo.
KF is supported by KAKENHI No.16H02211, JST PRESTO JPMJPR1668, JST ERATO JPMJER1601, and JST CREST JPMJCR1673. This work is supported by MEXT, Q-LEAP. 
\bibliography{bibliography}
\onecolumngrid
\appendix
\section{Derivation of \cref{eq:cost_1p}}

Using \cref{eq:rotation_gate_condition}, the $j$-th rotation gate $R_j(\theta_j)$ can be rewritten as
\begin{equation}
    R_j(\theta_j) = I\cos\frac{\theta_j}{2} - iA_j\sin\frac{\theta_j}{2}.
\end{equation}

The circuit $U_j^{(n)}(\theta_j)$ can be divided into three parts as follows:
\begin{equation}
    U_j^{(n)}(\theta_j) = V_{1j}^{(n)}R_j(\theta_j)V_{2j}^{(n)},
\end{equation}
where $V_{1j}^{(n)}$ and $V_{2j}^{(n)}$ are unitary operators which are independent of $\theta_j$.
The cost function $\mathcal{L}_j^{(n)}(\theta_j)$ can then be transformed as
\begin{align}
    \begin{split}
    \mathcal{L}_j^{(n)}(\theta_j)
    =& \sum_{k=1}^K w_k
    \braket{\varphi_k|
    U_j^{(n)\dagger}(\theta_j)\mathcal{H}_k U_j^{(n)}(\theta_j)
    |\varphi_k}
    \\
    =& \sum_{k=1}^K w_k
    \braket{\varphi_k|
    V_{2j}^{(n)\dagger}R_j(-\theta_j)V_{1j}^{(n)\dagger}
    \mathcal{H}_k
    V_{1j}^{(n)}R_j(\theta_j)V_{2j}^{(n)}
    |\varphi_k}
    \\
    =& \sum_{k=1}^K w_k
    \braket{{\varphi'}_{kj}^{(n)}|
    \qty(I\cos\frac{\theta_j}{2} + iA_j\sin\frac{\theta_j}{2})
    \mathcal{H'}_{kj}^{(n)}
    \qty(I\cos\frac{\theta_j}{2} - iA_j\sin\frac{\theta_j}{2})
    |{\varphi'}_{kj}^{(n)}}
    \\
    &\hspace{10mm} \qty(
    \ket{{\varphi'}_{kj}^{(n)}} := V_{2j}^{(n)}\!\ket{\varphi_k},\quad
    \mathcal{H'}_{kj}^{(n)} := V_{1j}^{(n)\dagger}\mathcal{H}_k V_{1j}^{(n)}
    )
    \\
    =& \cos^2\frac{\theta_j}{2} \sum_{k=1}^K w_k
    \braket{{\varphi'}_{kj}^{(n)}|
    \mathcal{H'}_{kj}^{(n)}
    |{\varphi'}_{kj}^{(n)}}
    \\
    &+ \sin^2\frac{\theta_j}{2} \sum_{k=1}^K w_k
    \braket{{\varphi'}_{kj}^{(n)}|
    A_j \mathcal{H'}_{kj}^{(n)} A_j
    |{\varphi'}_{kj}^{(n)}}
    \\
    &+ \sin\frac{\theta_j}{2}\cos\frac{\theta_j}{2} \sum_{k=1}^K w_k
    \braket{{\varphi'}_{kj}^{(n)}|
    i\qty(A_j \mathcal{H'}_{kj}^{(n)} - \mathcal{H'}_{kj}^{(n)} A_j)
    |{\varphi'}_{kj}^{(n)}}
    \\
    =& c_{1j}^{(n)}\cos^2\frac{\theta_j}{2} + c_{2j}^{(n)}\sin^2\frac{\theta_j}{2} + c_{3j}^{(n)}\sin\frac{\theta_j}{2}\cos\frac{\theta_j}{2}
    \quad (c_{1j}^{(n)}, c_{2j}^{(n)}, c_{3j}^{(n)} \in \mathbb{R})
    \\
    =& c_{4j}^{(n)}\cos\theta_j + c_{5j}^{(n)}\sin\theta_j + c_{6j}^{(n)}
    \quad (c_{4j}^{(n)}, c_{5j}^{(n)}, c_{6j}^{(n)} \in \mathbb{R})
    \\
    =& a_{1j}^{(n)} \cos(\theta_j - a_{2j}^{(n)}) + a_{3j}^{(n)}
    \quad (a_{1j}^{(n)}, a_{2j}^{(n)}, a_{3j}^{(n)} \in \mathbb{R}),
    \label{eq:deriv_normal}
    \end{split}
\end{align}
where $c_{\ell j}^{(n)} \  (\ell=1,2,\cdots,6)$ and $a_{\ell j}^{(n)} \  (\ell=1,2,3)$ denote real constants independent of $\theta_j$.

\section{Derivation of \cref{eq:cost_general}}
Mathematical induction can be used to prove that the following statement, $P(m)$, holds for all natural numbers $m$.
\begin{equation}
    P(m): \ \exists\bm{b} \quad
    \braket{\varphi|
    U_m^\dagger\qty(\qty{\theta_j}_{j=1}^m)
    \mathcal{H}
    U_m\qty(\qty{\theta_j}_{j=1}^m)
    |\varphi}
    = \bm{b} \cdot \qty[ \bigotimes_{j=1}^m
    \mqty(\cos\theta_j \\ \sin\theta_j \\ 1)],\label{eq:statement_general}
\end{equation}
where $\bigotimes$ denotes Kronecker product,
$U_m(\qty{\theta_j}_{j=1}^m)$ is a given parameterized quantum circuit which has $m$ parameters and which satisfies the three conditions of \cref{sec:preconditions},
$\mathcal{H}$ is a given Hamiltonian,
and $\ket{\varphi}$ is a quantum state.
The order of the parameters $\qty{\theta_j}_{j=1}^m$ is the same as the order of operating corresponding rotation gates.

When $m=1$, the statement can prove in the similar manner as \cref{eq:deriv_normal}.

Show that if $P(t)$ holds, then also $P(t+1)$ holds.
This can be done as follows.

Assume $P(t)$ holds for some unspecified value of $t$.
It must then be shown that $P(t+1)$ holds as follows.

The given circuit $U_{t+1}\qty(\qty{\theta_j}_{j=1}^{t+1})$ can be divided into three parts as follows:
\begin{equation}
    U_{t+1}\qty(\qty{\theta_j}_{j=1}^{t+1})
    = U_0 R_{t+1}\qty(\theta_{t+1}) U_t\qty(\qty{\theta_j}_{j=1}^{t}),
\end{equation}
where $U_0$ denotes a unitary with no parameter, and $U_t\qty(\qty{\theta_j}_{j=1}^{t})$ denotes a unitary with $t$ parameter.

Therefore,
\begin{align}
    \begin{split}
    &\hspace*{-.5cm}\braket{\varphi|
    U_{t+1}^\dagger\qty(\qty{\theta_j}_{j=1}^{t+1})
    \mathcal{H}
    U_{t+1}\qty(\qty{\theta_j}_{j=1}^{t+1})
    |\varphi}
    \\
    =&\braket{\varphi|
    U_t^\dagger\qty(\qty{\theta_j}_{j=1}^{t}) R_{t+1}(-\theta_{t+1}) U_0^\dagger
    \mathcal{H}
    U_0 R_{t+1}(\theta_{t+1}) U_t\qty(\qty{\theta_j}_{j=1}^{t})
    |\varphi}
    \\
    =&\braket{\varphi|
    U_t^\dagger\qty(\qty{\theta_j}_{j=1}^{t})
    \qty(I\cos\frac{\theta_{t+1}}{2} + iA_{t+1}\sin\frac{\theta_{t+1}}{2})
    \mathcal{H'}
    \qty(I\cos\frac{\theta_{t+1}}{2} - iA_{t+1}\sin\frac{\theta_{t+1}}{2})
    U_t\qty(\qty{\theta_j}_{j=1}^{t})
    |\varphi}
    \\
    &\hspace{10mm} \qty(
    \mathcal{H'} := U_0^\dagger \mathcal{H} U_0
    )
    \\
    =&\cos^2\frac{\theta_{t+1}}{2}
    \braket{\varphi|
    U_t^\dagger\qty(\qty{\theta_j}_{j=1}^{t})
    \mathcal{H'}
    U_t\qty(\qty{\theta_j}_{j=1}^{t})
    |\varphi}
    \\
    &+\sin^2\frac{\theta_{t+1}}{2}
    \braket{\varphi|
    U_t^\dagger\qty(\qty{\theta_j}_{j=1}^{t})
    A_{t+1}\mathcal{H'}A_{t+1}
    U_t\qty(\qty{\theta_j}_{j=1}^{t})
    |\varphi}
    \\
    &+\sin\frac{\theta_{t+1}}{2}\cos\frac{\theta_{t+1}}{2}
    \braket{\varphi|
    U_2^\dagger\qty(\qty{\theta_j}_{j=1}^t)
    i\qty(A_{t+1}\mathcal{H'} - \mathcal{H'}A_{t+1})
    U_2\qty(\qty{\theta_j}_{j=1}^t)
    |\varphi}
    \\
    =&\bm{b}_1 \cdot \qty[ \bigotimes_{j=1}^t
    \mqty(\cos\theta_j \\ \sin\theta_j \\ 1)]
    \cos^2\frac{\theta_{t+1}}{2}
    \\
    &+\bm{b}_2 \cdot \qty[ \bigotimes_{j=1}^t
    \mqty(\cos\theta_j \\ \sin\theta_j \\ 1)]
    \sin^2\frac{\theta_{t+1}}{2}
    \\
    &+\bm{b}_3 \cdot \qty[ \bigotimes_{j=1}^t
    \mqty(\cos\theta_j \\ \sin\theta_j \\ 1)]
    \sin\frac{\theta_{t+1}}{2}\cos\frac{\theta_{t+1}}{2}
    \quad \qty(\bm{b}_1,\bm{b}_2,\bm{b}_3\in{\mathbb{R}^{3^t}})
    \\
    =&\bm{b}_4 \cdot \qty[ \bigotimes_{j=1}^t
    \mqty(\cos\theta_j \\ \sin\theta_j \\ 1)]
    \cos\theta_{t+1}
    \\
    &+\bm{b}_5 \cdot \qty[ \bigotimes_{j=1}^t
    \mqty(\cos\theta_j \\ \sin\theta_j \\ 1)]
    \sin\theta_{t+1}
    \\
    &+\bm{b}_6 \cdot \qty[ \bigotimes_{j=1}^t
    \mqty(\cos\theta_j \\ \sin\theta_j \\ 1)]
    \quad \qty(\bm{b}_4,\bm{b}_5,\bm{b}_6\in{\mathbb{R}^{3^t}})
    \\\
    =&\bm{b} \cdot \qty[ \bigotimes_{j=1}^{t+1}
    \mqty(\cos\theta_j \\ \sin\theta_j \\ 1)]
    \quad \qty(\bm{b}\in{\mathbb{R}^{3^{t+1}}}).\label{eq:deriv_general}
    \end{split}
\end{align}
Thereby showing that indeed $P(t+1)$ holds.

Since both the base case and the inductive step have been performed,
by mathematical induction the statement $P(m)$ holds for all natural numbers $m$.

%Using \cref{eq:statement_general}, we can derive \cref{eq:cost_general}.

\section{Derivation of \cref{eq:cost_special}}

Assume that the parameterized quantum circuit has $S$ rotation gates which has the same parameter $\theta$.
Because of \cref{eq:deriv_general}, then cost function can written by
\begin{equation}
    % \bm{b} \cdot \qty[ \bigotimes_{s=1}^S
    % \mqty(\cos\theta_s \\ \sin\theta_s \\ 1)].
    \bm{b} \cdot \mqty(\cos\theta \\ \sin\theta \\ 1)^{\!\!\otimes S}.
\end{equation}
% Adding to this the constraint which the $S$ parameters are the same parameter $\theta$, 
This equation is written as follows:
\begin{align}
    \begin{split}
    % \bm{b} \cdot \qty[ \bigotimes_{s=1}^S
    % \mqty(\cos\theta \\ \sin\theta \\ 1)]
    \bm{b} \cdot \mqty(\cos\theta \\ \sin\theta \\ 1)^{\!\!\otimes S}
    % \quad \qty(\bm{b}\in{\mathbb{R}^{3^N}})
    =& \sum_{\smqty{p,q\in\mathbb{N}\\p+q\leq S}}
    a_{p,q} \cos^p\theta \sin^q\theta
    \quad(a_{p,q} \in \mathrm{R})
    \\
    =& \sum_{s=0}^S \eta_s \cos^s\theta
    + \sum_{s=0}^{S-1} \xi_s \cos^s\theta \sin\theta
    \quad(\eta_s,\xi_s \in \mathbb{R})
    \\
    =& \sum_{s=0}^S \eta'_s \cos(s\theta)
    + \sum_{s=0}^{S-1} \xi'_s \cos(s\theta) \sin\theta
    \quad(\eta'_s,\xi'_s \in \mathbb{R})
    \\
    =& \sum_{s=0}^S \eta'_s \cos(s\theta)
    + \sum_{s=0}^{S-1} \frac{\xi'_s}{2} \qty(\sin((s+1)\theta) - \sin((s-1)\theta))
    \\
    =& \sum_{s=0}^S \eta'_s \cos(s\theta)
    + \sum_{s=0}^S \xi''_s \sin(s\theta)
    \quad(\xi''_s \in \mathbb{R})
    \\
    =& \sum_{s=1}^S a_s \cos(s\theta) + \sum_{s=1}^S b_s \sin(s\theta) + c
    \quad(a_s,b_s,c \in \mathbb{R}).
    \end{split}
    \label{eq:deriv_special}
\end{align}

Using
Eq.~(C2), %\cref{eq:deriv_special},  % bug of TeX?
we can derive \cref{eq:cost_special}.

\end{document}